# Dosimetric characterization of a new $^{192}$Ir pulse dose rate brachytherapy source with the Monte Carlo simulation and thermoluminescent dosimeter

**Vahid. Lohrabian** [1], **Alireza Kamali-Asl** [1], **Hossein Arabi**[2], **Hamidreza Hemmati** [1], **Majid Pournezam Esfahani** [3]

[1]Department of Medical Radiation Engineering, University of Shahid Beheshti, Tehran, Iran

[2]Division of Nuclear Medicine and Molecular Imaging, Department of Medical Imaging, Geneva University Hospital, CH-1211 Geneva 4, Switzerland

[3]Department of Biomedical Engineering, Ajman University, Al Jerf 1, PO Box 346, Ajman, United Arab Emirates.

**Abstract**

**Introduction:** Pulse dose rate (PDR) brachytherapy is a highly extended practice in clinical brachytherapy research these days. In general, important parameters regarding the dose distribution of radioactive sources used in clinical practice should be accurately calculated and characterized. Due to the different source designs, a specific dosimetry dataset is required for each source model. Monte Carlo calculations are highly spread and established practice to calculate brachytherapy sources by dosimetric parameters. In this study, recommendations of the AAPM TG- 43 (U1) report have been followed to characterize the new $^{192}$Ir pulse dose rate source, provided by the Applied Radiation Research School, Nuclear Science and Technology Research Institute in Iran.

**Material & Methods:** Dose rate constant, radial dose function, geometry factors and anisotropy function were calculated according to the relevant American Association of Physicists in Medicine AAPM and TG43 (U1) reports. In this study, $^{192}$Ir source was characterized using Monte Carlo simulation in water phantom, and in addition, experimental measurements were carried out using thermoluminescent dosimeters (TLD-100) in plexiglass (PMMA) phantoms.

**Results:** The dose-rate constant for the $^{192}$Ir PDR was found to be equal 1.131±0.007 cGyh$^{-1}$U$^{-1}$ and 1.173±0.005 cGyh$^{-1}$U$^{-1}$ with TLD measurement and Monte Carlo simulation, respectively. Also in this study, the geometry function $G(r,\theta)$, radial dose functions $g(r)$, and the anisotropy function $F(r,\theta)$ have been calculated at distances from 0.1 to 16 cm. The dose-rate constant of these calculations has been compared with measured values for an actual $^{192}$Ir seed. The results of dosimetry parameters, presented in tabulated and graphical formats, exhibited good agreement to those reported from other commercially available PDR $^{192}$Ir sources.

**Conclusions:** The results obtained in this study are in close agreement with the characteristics of the commercially available $^{192}$Ir sources. The results obtained in this study can be treated as an initial assessment of this source to be employed in the conventional treatment planning systems subsequent to complementary investigations.

# 1. Introduction

$^{192}$Ir, $^{137}$Cs, $^{125}$I and $^{103}$Pd Brachytherapy sources are widely used for interstitial implants in various tumor sites. For example, low-energy $^{125}$I sources, whose emission diminishes sharply with increasing distance from the source, are used for prostate tumors [1]. Appropriate treatments have been reported in this area by the permanent implantation of these sources in various hospitals around the world. Generally, these radioactive sources have been massively used for the treatment of cervical, brain, breast, penis, and prostate cancers as well as rectal tumors; and the results reported by specialists confirmed the effective impact of these commercial sources [2].

Brachytherapy sources containing $^{192}$Ir have been frequently used for permanent implants of the brain and other anatomic sites. In addition to the strong inverse-square law reduction in gamma ray fluence at short distances, photon emissions of $^{192}$Ir PDR sources lead to a rapid decrease in radiation dose with increasing distance. Therefore, this effect reduces the unnecessary radiation dose to normal tissue located beyond the tumor [3]. Recently, a new design of $^{192}$Ir PDR brachytherapy seed has been produced for clinical applications by the Applied Radiation Research School, Nuclear Science and Technology Research Institute, Tehran, Iran. According to American Association of Physicists in Medicine [4] TG-43(U1) [5] recommendations, before using each new source clinically, the dosimetric characteristics of the source must be determined in order to provide reliable data for treatment planning calculations and dose prescription [6]. The AAPM has recommended that dosimetry characteristics of all such sources should be established by two independent investigators, theoretical calculations and experimental measurements.

Pulsed dose rate (PDR) is a new modality for dose delivery in brachytherapy. It uses modern after-loading technology (miniaturized source, cable driven, software controlled), with source activities in the range of 1 Ci, which is actually one tenth of the normal activity used for high dose rate (HDR) brachytherapy [7, 8]. The process of delivering pulsed-dose-rate (PDR) brachytherapy is similar to HDR; however, radiation is delivered in short 'pulses' over several hours [9, 10]. The principle behind PDR is its biological similarity to traditional low dose rate (LDR) techniques in which sources are manually loaded into catheters and applicators. The source, then, is left in place for a number of hours or days. Since PDR uses a remote controlled after-loader, it promotes radiation safety, prospective planning, and precise dose shaping [9, 11]. In contrast to the HDR technique, PDR treatments are delivered on an in-patient basis and require a dedicated shielded treatment room where a patient can stay for up to one day. Pulsed dose rate (PDR) was developed, adapting the mechanics of modern HDR after-loader. As in HDR brachytherapy, PDR makes use of a single source (usually $^{192}$Ir) of high activity (usually between 0.5 and 1 Ci), secured at the end of a cable-driven wire. Contrary to HDR treatment, where large doses per fraction are delivered (usually in the range of 5 Gy), the peculiarity of the PDR concept involves the delivery of the brachytherapy dose in a large number of small fractions (pulses) over several days in order to approximate the radiobiological characteristics of low dose rate (LDR) irradiation. Some theoretical studies determined biological equivalence between PDR and conventional, gold standard, continuous LDR; and initial protocols set the basis for its routine use in the clinical setting [12]. Therefore, PDR after-loader has been proposed as a method of replacing continuous LDR, assuming radiobiological equivalence [9, 10, 13]. It is recommended that a realistic geometry with respect to the physical properties and radioactivity characteristics of the source should be implemented in experimental test and/or Monte Carlo (MC) simulation to properly define the inputs to the treatment frameworks using HDR and PDR sources [6]. In clinical trials, choosing between HDR and PDR techniques is rather possible; however, designing specific treatments is not easily distinguishable.

Current treatment planning systems (TPS) used for high-dose-rate HDR, pulsed-dose-rate(PDR), and low-dose rate (LDR) brachytherapy allow direct introduction of tabulated dose rates from the literature using the TG-43(U1) formalism [5]. These TG-43(U1) data are usually derived from Monte Carlo radiation transport simulations for the estimation of absorbed dose by collision kerma, $S_K$. Consequently, these data are provided at distances from the source capsule large enough to assure the validity of the equivalence of kerma and dose. TPS extrapolates data outside the available TG-43(U1) data range [9, 10, 13].

The TG-43(U1) dose-rate constant, $\Lambda$, for this source is based on thermoluminescent detector (TLD) dose-rate measurements in a Plexiglass phantom (PMMA) which were normalized to the nominal air-kerma strength ($S_K$), derived from the vendor's in-house apparent activity ($A_{app}$) standard [5, 14, 15].

Casado et al. calculated dose rate constant as well as radial dose function using PENELOPE simulation, which is consistent with our results [16]. Moreover, Sarabiasl also reported the value of dose rate constant for HDR Ir-192 source with MCNP simulation, indicating a good agreement with the source dose rate constant used in this article [17]. Recently, Badry et al, exploited EGS5 simulation and introduced HDR Ir-192 brachytherapy dose distribution parameters, whose research results are in a good agreement with the results of the present research.[18]. In a similar study, Taylor et al. employed EGSnrc Monte Carlo calculation to obtain the dose distribution surrounding a high dose rate $^{169}$Yb brachytherapy source as well as 14 high dose rates and pulsed dose rates for $^{192}$Ir brachytherapy sources [19].

In this study, carried out similarly to the previous research, some modifications have been made in order to simulate clinical practice conditions in the testing condition primarily due to the increasing importance of using these radioactive sources. The modifications include the differences in the source's dimensions, radioactivity, type of detectors and the material of the phantoms used.

The goal of this project was the experimental and theoretical determination of the dosimetric characteristics of the $^{192}$Ir PDR source. This investigation was performed using LiF TLD chips and Monte Carlo simulations following the recommendations in the AAPM TG-43(U1) protocol [5]. Several dosimetric parameters of this source such as dose-rate constant, radial dose function, 2D anisotropy functions as well as 2D along-away rectangular dose rate lookup tables are calculated. The Monte Carlo calculation code and experimental results used in this study have been compared to measurement results for the same types of this source seed [20].

## 2. Materials and Methods

### 2.1. $^{192}$Ir PDR source

Fig. 1 shows a schematic diagram of the $^{192}$Ir PDR source. The source has active length of 3 mm, an outer diameter of 0.6 mm. The interior fountain configuration includes a thickness of about 0.15 mm, a combination of $^{192}$Ir and Pt metals, surrounded by Platinum metal. The cladding is intended to absorb Beta rays. The activity of the source used in this study is 140 cGy. A metal coating was used to absorb beta radiation. The photon energy spectrum of the $^{192}$Ir source is shown in table 1 in accordance with the TG-43(U1), [5] recommendation and physical characteristics of the material used in this brachytherapy source are also listed in table 2 [21, 22].

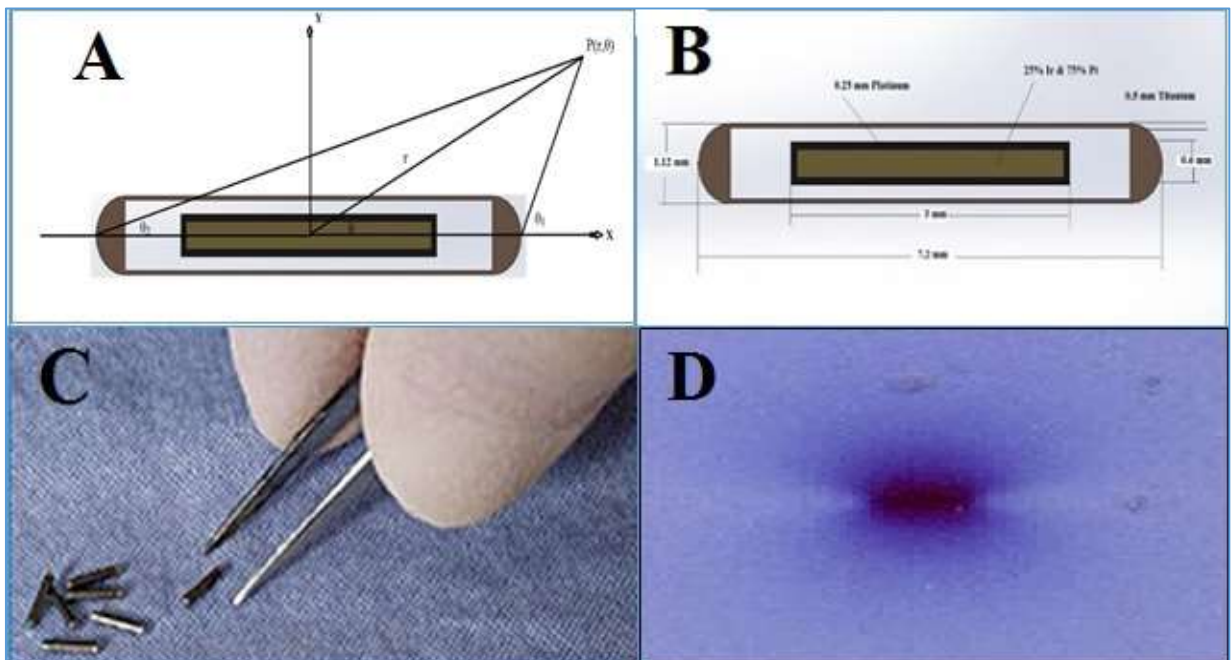

**Figure 1.** A) Diagram of geometry assumed in TG43 (U1). The distance (r) and the dosimetric point of the source from the point of view of the source ($\theta_1$ & $\theta_2$) are specified. B) Schematic diagram of the PDR Ir-192 seed and its materials. C) Real radioactive seed source. D) Radioactive source traces on Gafchromic film (EBT).

Table 1. Photon energy spectrum of the $^{192}$Ir source.

| Photon energy (keV) | 0.296 | 0.308 | 0.317 | 0.468 | 0.485 | 0.589 | 0.604 | 0.612 |
|---|---|---|---|---|---|---|---|---|
| Collapse for each decay | 0.290 | 0.297 | 0.828 | 0.478 | 0.0316 | 0.0452 | 0.0818 | 0.0533 |

Table 2. Physical characteristics of the elements forming the source PDR-Ir192.

| Material | Atomic Number[23] | Mass Density(g/cm$^3$) | Simulated physical state | Composition |
|---|---|---|---|---|
| Platinum | 78 | 21.45 | Solid | 25% |
| Titanium | 22 | 4.506 | Solid | 100% |
| Iridium | 77 | 22.56 | Solid | 75% |
| Air | 7.62 | 0.00120 | Gas | 100% |

The active Iridium core has been produced from the neutron activation by the 5 MW Tehran Research Reactor (TRR) at NSTRI (Nuclear Science and Technology Research Institute, Tehran, Iran). The half-life of this source is 74.7 days and its gamma-ray spectrum is in the energy range between 0.14 MeV and 1.06 MeV and its average energy is 0.36 MeV. This moderate amount of energy makes it unnecessary to use large shields when working with this source. The $^{192}$Ir source method used in this research is according to the following nuclear reaction [11, 21]:

$$^{191}_{77}\text{Ir} + ^{1}_{0}\text{n} \rightarrow \gamma + ^{192}_{77}\text{Ir} \tag{1}$$

## 2.2. TLD dose measurements

Dose distributions around the $^{192}$Ir source were measured in plexiglass (Polymethyl Methacrylate) or PMMA phantom using TLD-100 (LiF:Mg) chips. The TLD chips with dimensions of 3.1×3.1×0.9 mm$^3$ were used in all phantom measurements evaluating the $^{192}$Ir source, and in Lucite build-up capsules using the $^{60}$Co source. An automated TLD reader was used (Harshaw-Bicron, model 4500) and chips were placed in the phantom in a special pattern to minimize inter-rod effects. The following section is a brief summary of these procedures.

The irradiated TLDs were read using a Harshaw Model 4500 TLD reader and were annealed using a standard technique. The following equation was used to calculate the absorbed dose rate from the TLD responses from each point irradiated in the phantom [14, 22, 24]:

$$\frac{\dot{D}(r,\theta)}{S_K} = \frac{R}{TS_K \varepsilon E(r)d(T)F_{lin}} \tag{2}$$

Where R is the TLD response, corrected for the physical differences of the TLD chips using the predetermined chip factors. $\dot{D}(r,\theta)$ is the dose rate at the r distance and the specific theta angle. T is the experimental time period (hours) and $S_K$ is the measured source strength at the initial time of measurement. $\varepsilon$ is the calibration factor for the TLD response (nC/cGy) measured with a 6 MV x-ray beam from a linear accelerator. $E(r)$ is the correction factor for the energy dependence of the TLD between the calibration beam and the $^{192}$Ir photons. A value of 1.4 was used for $E(r)$ in this project. $d(T)$ is a correction for the source decay during the experimental time period. $F_{lin}$ is the nonlinearity correction of the TLD response for the given dose [24-26]. In measurements, the experimental times were selected such that the absorbed doses ranged from 10–100 cGy, the range over which the TLD response is linear.

To perform the dosimetry and its associated calibration, the choice of phantom type used to create maximum dispersion is very important. Hence, the effective atomic number parameter, which is equivalent to the texture, is very important because of its capability to create the same conditions as the real test. Water, Acrylics, Polystyrene or similar low atomic number materials are appropriate for such phantoms [14, 21]. Polymethyl Methacrylate (PMMA) or Plexiglass is one of such materials with low effective atomic number ($Z_{eff}$=6.5) suitable for calibration and dose measurements.

Fig. 2a-d shows the schematic diagram of the experimental setup measurement of dose rate constant, radial dose function and anisotropy function. The TLD data at each point was taken as the average values from the four measurements with an uncertainty of about 7% of standard deviation [5].

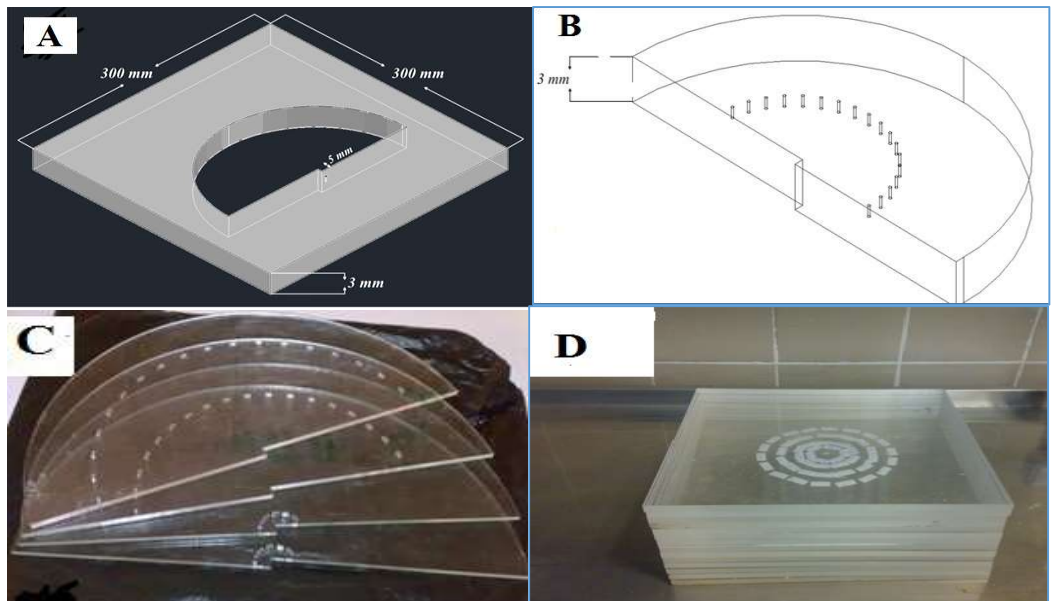

**Figure 2.** A) 3D sketch of the host of the Plexiglass Phantom. (B) Simulate the location of TLDs. (C) Plexiglass phantom. (D) Phantom and source under test conditions.

## 2.3. Monte Carlo Simulation

The MCNP Monte Carlo code is a powerful modeling tool in the simulation and transporting of different particles and radiation the various geometries. In this study, The MCNP Monte Carlo code was employed to simulate the solid 100% water phantom, $^{192}$Ir gamma-ray brachytherapy source and several scoring cells for calculating the absorbed dose at different distances from the source. It should be mentioned that all of the required data for simulation have been taken from previous researches [27].

The gamma source characteristics including the effective energy at the $^{192}$Ir nominal energies and intensity profile of incident photons has been regarded in the simulations. The energy cut off of photon beam 0.01 MeV was considered. Two and three-dimensional view of simulated setup for $^{192}$Ir gamma source are shown in Fig. 3 [7, 28, 29]. The size of scoring cells for dose calculation was set to 0.3×0.25×0.25 cm$^3$ and for measuring the absorbed dose along the central axis of the photon beam into 25×30×30 cm$^3$, water phantom dimensions was simulated. It should be emphasized that the medium of scoring cells has been simulated with water material. The number of followed histories in each simulation was set to 9×10$^9$. The absorbed dose distribution at different distances of $^{192}$Ir brachytherapy source is shown in terms of Gy. The statistical uncertainty in all of the MC simulations was less than 3%.

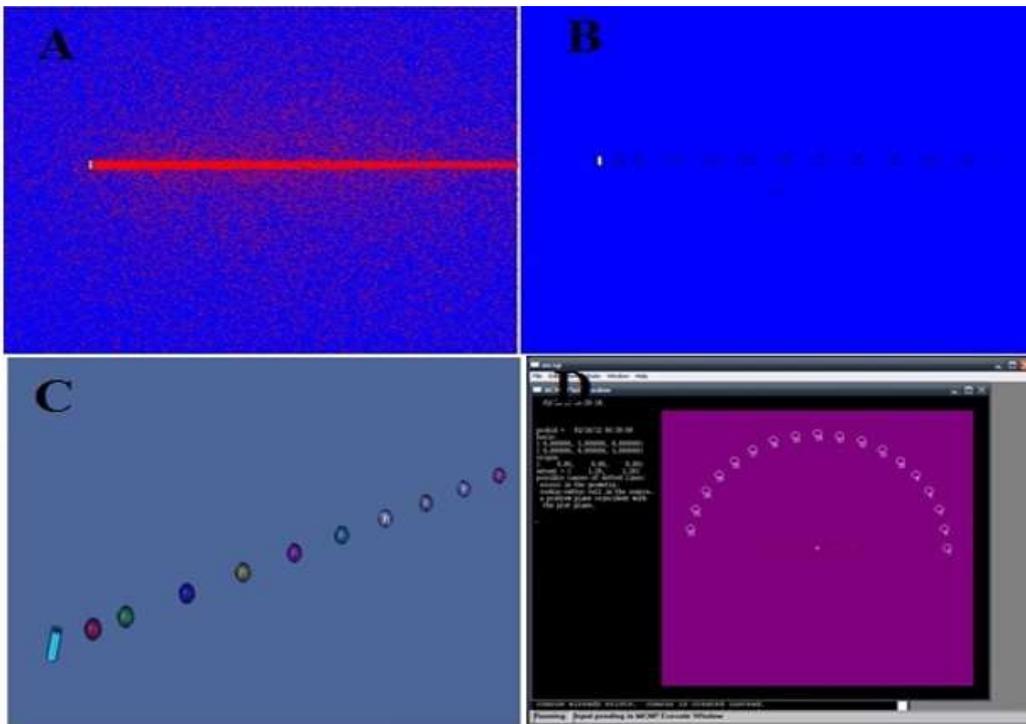

**Figure 3.** A) Particle transport in simulation environment. B) Layout of dosimetric points in the main axis of the source C) Layout of dosimetric points in the direction perpendicular to the main axis of the source. D) Layout of dosimetric points at a specific radial distance from the source.

In general, many standard tallies are used by the MCNP code to determine the mentioned parameters [30, 31]. In this study, *F8 tally (deposited energy) was used for close distances to the source, while the F6 tally (track length estimate of energy deposition) was employed for far distances due to the existence of the electronic equilibrium condition. It should be mentioned that the updated ENDF/B-VI.8 library which is based on the EPDL97 library [27] was used to allow compatible and consistent transport of photon, electron and coupled photon-electron. Note that EPDL97 data are available online in ENDF format as part of ENDF/B-VI Atomic data [32]. For validation of MC simulation, the absorbed dose at several distances was measured by Thermoluminescence dosimetry. The gamma analysis was employed for the comparison of MC simulation and TLD measurements. The gamma index calculations were performed in Gnuplot (version 4.4 patch level 3; Geeknet Inc, Fairfax, VA, USA) software environment using gamma index executive file. The dose difference and distance to agreement in the calculation of gamma index were considered as 3% and 3 mm, respectively (as recommended by Low et al [33]). In this work, $9 \times 10^9$ photons histories were used for the simulations of $g_L(r)$ and $F(r, \theta)$ to obtain statistical uncertainties below 2% as described in TG-43(U1) [5] inside a spherical water phantom with a radius of 50 cm. The conditions for the location of the radioactive source in the simulation environment along with the corresponding distances are shown in Figs, 3.

For the experimental measurement, the radioactive sources were placed inside the plexiglass phantom. As such, for consistency in the simulation, the same source dimensions were implemented in water (equivalent to the soft-tissue).

## 2.4. Dosimetric calculations

The radioactive source is placed in the center of a spherical phantom that is 50 cm in diameter for calculations in water according to TG43(U1). Since in laboratory condition, the phantom and the source were in the air environment, the simulated phantom and the source were placed in a sphere about 5 meters in diameter filled with air. The composition of materials used in this work is shown in Table 2. TG-43(U1) formalism sets up a general expression for absorbed dose rate in water around brachytherapy sources [5]:

$$\dot{D}(r,\theta) = S_K \Lambda \frac{G(r,\theta)}{G(r_0,\theta_0)} g(r) F(r,\theta) \qquad (3)$$

Dose rate constant ($\Lambda$) represents the dose rate to the medium (Plexiglass) at the reference point $r = 1cm$ along the transverse axis $\theta = \pi/2$ of the source per air kerma strength.

$$\Lambda = \frac{\dot{D}(1, \pi/2)}{S_K} \qquad (4)$$

Having $S_K$ and determining the dose rate at 1 cm distance from the source center along its transverse axis ($\dot{D}(1, \pi/2)$), the dose rate constant was calculated using Eq.4 [5]. Air kerma strength is a measure of brachytherapy source strength, which is specified in terms of air kerma rate at points located at 1 m distance along the transverse axis of the source in free space. With regards to the units of µGy, h and m, for kerma, time and distance, respectively. The unit of $S_k$ will be µGym² h⁻¹ as recommended in TG-43 (U1) report, which is denoted by symbol of U [5]:

$$1U = 1\mu Gy m^2 h^{-1} = 1 cGy cm^2 h^{-1} \tag{5}$$

To determine the dose rate constant, first the source air kerma rate $\dot{K}(d)$ was calculated within an air-filled spherical scoring cell with radius of 4 mm located at 50 cm distance from the source center along its transverse axis. It will be followed according the procedure described by Williamson et al [34], calculating air kerma rate in the air, from 2 to 50 cm away from the source, every 1 cm along the Y-axis. Air kerma strength was calculated using the following Equation [27, 29]:

$$\dot{K}(d)d^2 = S_K + \alpha d \quad \& \quad d = 50 cm \tag{6}$$

Where $\alpha$ describes the deviation. Due to the attenuation and scattering of photons in the air of air kerma rate with regards to inverse square law, the intervening medium between the source and the air-filled scoring cell was considered as a vacuum. It should be mentioned that the characteristic X-rays emitted from source clad and other low energy X-rays were also excluded from calculations by setting particle cutoff energy to 5keV, as recommended by TG-43(U1) protocol [5, 26, 27, 35].

One of the important parameters in dosimetry is the function of the source geometry. This parameter considers the effects photon flux reduction, happening as a result of the distance from the source, and depends on the distribution of radioactive material as well as the source nucleus (16, 22). This parameter can be defined as follows [5]:

$$G(r,\theta) = \frac{\tan^{-1}\left[\frac{r\cos\theta + L/2}{r\sin\theta}\right] - \tan^{-1}\left[\frac{r\cos\theta - L/2}{r\sin\theta}\right]}{Lr\sin\theta} \tag{7}$$

Where L is the active length of the source which is considered as 3 mm.

Radial dose function, $g(r)$, describes the attenuation of the photons emitted from the brachytherapy source in tissue. To determine the radial dose function, g(r), scoring cells with 0.5 mm radius were located at 0.1 to 10 cm distance from the source center along its transverse axis. The radial dose function is defined as [5]:

$$g(r) = \frac{\dot{D}(r, \frac{\pi}{2}) G(r_0, \frac{\pi}{2})}{\dot{D}(1, \frac{\pi}{2}) G(r, \frac{\pi}{2})} \tag{8}$$

where $\dot{D}(r, \frac{\pi}{2})$ and $\dot{D}(1, \frac{\pi}{2})$ are the measured dose rates, $G(1, \frac{\pi}{2})$ and $G(r, \frac{\pi}{2})$ are the geometry functions at distances of 1 cm and $r$ cm, respectively, along the transverse axis of the source. The relative difference (RD%) for the values of the radial dose function is obtained from the equation below and by reference data.

$$RD\% = \frac{|Value_{Thiswork} - Value_{refrence}|}{Value_{refrence}} \tag{9}$$

Anisotropy function is the dose changes around the source due to the distribution of radioactivity in it and the absorption and dispersion of photons in the phantom environment. In fact, the structural symmetry of the source shows a 90-degree polar angle and the same radial distances. The anisotropy function is defined as:

$$F(r, \theta) = \frac{\dot{D}(r, \theta) G(r, \frac{\pi}{2})}{\dot{D}(r, \frac{\pi}{2}) G(r, \theta)} \tag{10}$$

Where $\dot{D}(r, \theta)$ and $\dot{D}(r, \frac{\pi}{2})$ are the dose rates measured at the distance of $r$ cm and angles of $\theta$ and $\pi/2$ relative to the longitudinal axis of the source, respectively. The anisotropy factor is defined following the TG-43(U1) recommendations as [5]:

$$\phi_{an}(r) = \frac{\int \dot{D}(r, \theta) \sin \theta \, d\theta}{2 \dot{D}(r, \frac{\pi}{2})} \tag{11}$$

The anisotropy constant, $\varphi_{an}$, of the new source was determined by averaging the individual anisotropy factors in the given medium.

The energy spectrum of $^{192}$Ir gamma rays was obtained using the data reported by NCRP 58 [36]. Also, according to the recommendations of TG-43(U1) protocol [5], humidity of the air between the cylindrical core and the source clad was considered to be 40% (Hydrogen: 0.9186%, Carbon: 1.2158%, Nitrogen: 73.0411%, Oxygen: 23.5231%, and Argon: 1.3014%, (data are in percent mass)) in all of the simulations.

## 3. Results

The dose rate constant (Λ) value at the reference point ($\dot{D}(1, \pi/2)$) in the phantom, equivalent to the texture with simulation and practicality, is calculated as 1.173±0.005 and 1.131±0.007, respectively. Comparing these results shows a good agreement between simulation and practicality. The difference between the measured and calculated data was about 3.4%. Comparison of the dose rate constant of $^{192}$Ir source to that of other commercial sources is shown in Table 3. As can be seen, there is a good agreement between the simulated dose rate constant of the $^{192}$Ir and that of other commercially available sources.

Table 3. Comparison of dose rate constant in PDR-Ir192 with other commercial sources.

| Source | Method | Medium | Dose rate constant(cGyh$^{-1}$U$^{-1}$) | Type |
|---|---|---|---|---|
| Meigooni[37] | Monte Carlo | Liquid Water | 1.128±0.5% | PDR |
| Williamson[34] | Monte Carlo | Liquid Water | 1.110±0.2% | LDR |
| Karaiskos microSelectron v2[32] | Monte Carlo | Solid Water | 1.121±0.006 | PDR |
| Perez-Calatayud GammaMed 12i[38] | Monte Carlo | Water | 1.122±0.003 | PDR |
| Nucletron microSelectron v1[39] | Monte Carlo | Liquid Water | 1.124±0.006 | PDR |
| Perez-Calatayud GammaMed Plus[38] | Monte Carlo | Water | 1.122±0.003 | PDR |
| Buchler G0814[35] | Monte Carlo | Solid Water | 1.115±0.003 | HDR |
| Ghiassi-Nejad[3] | TLD | Plexiglass | 1.196±0.060(seed 5 mm) | LDR |
| Ghiassi-Nejad[3] | TLD | Plexiglass | 1.082±0.054(seed 10 mm) | PDR |
| GammaMed HDR 12i[8] | Monte Carlo | Water | 1.108±0.0030 () | HDR |
| BEBIG-HDR-GI192M11[29] | Monte Carlo | Water Phantom | 1.117±0.0040 | HDR |
| Flexisource Ir-192[11] | Monte Carlo | Water Phantom | 1.109±0.0110 | HDR |
| mHDR-v2r[40] | Monte Carlo | Water Phantom | 1.112±0.0008 | HDR |
| HDR 192Ir IRAsource[17] | Monte Carlo | Water Phantom | 1.112±0.0050 | PDR |
| Varian GammaMed Plus [18] | Monte Carlo | Solid Water | 1.113±0.4% | HDR |
| This Work | TLD | Plexiglass | 1.131±0.007 | PDR |
| This Work | Monte Carlo | Water Phantom | 1.173±0.005 | PDR |

The values of the simulated radial dose function of $^{192}$Ir at various distances from the source center are reported in Table 4. The comparison between the measured and simulated radial dose function is shown in Fig. 4. As can be seen in Fig. 4, the results of Monte Carlo simulation are in accordance with TLD measured data. The difference between the simulated and measured radial dose function are shown in the table 4.

Table 4. Comparison of radial dose between Monte Carlo simulation and TLD measurement. The fourth column of this table is the error between simulated and practical measurements. The fifth column is the relative error between simulated and practical measurements with respect to those of reference TG43 (U1).

| distance from source center, r(cm) | Simulated g(r), Water Phantom | Measured in Plexiglass with LiF TLD | Relative difference (%) | Relative difference (%) with reference |
|---|---|---|---|---|
| 0.1 | 0.975 | - |  | 0.6 |
| 0.2 | 0.981 | - |  | 0.8 |
| 0.3 | 0.986 | - |  | 0.6 |
| 0.4 | 0.992 | - |  | 0.3 |
| 0.5 | 0.994 | 1.003 | 0.8 | 0.2-0.7 |
| 0.6 | 0.999 | - |  | 0.2 |
| 0.7 | 1.003 | - |  | 0.7 |
| 0.8 | 1.002 | - |  | 0.4 |
| 0.9 | 1.012 | - |  | 1.3 |
| 1 | 1 | 1 | 0 | 0 |
| 1.5 | 0.978 | 0.974 | 0.6 | 2.5-2.7 |
| 2 | 0.999 | 0.989 | 1 | 0.7-1.6 |
| 2.5 | 1.003 | 0.992 | 1 | 0.1-1.3 |
| 3 | 1.018 | 1.003 | 1.4 | 1.2-0.2 |
| 3.5 | 0.981 | 0.986 | 0.5 | - |
| 4 | 0.983 | 0.995 | 1.2 | 2.2-0.1 |
| 4.5 | 0.974 | 0.972 | 0.2 | - |
| 5 | 1.011 | 0.997 | 1.3 | 0.9-0.5 |
| 5.5 | 0.972 | 0.967 | 0.5 | - |
| 6 | 0.966 | 0.966 | 0.7 | 3.1-3.1 |
| 6.5 | 0.963 | 0.953 | 3.1 | - |
| 7 | 0.954 | 0.956 | 0.3 | 3-2.9 |
| 7.5 | 0.937 | 0.937 | 1 | - |
| 8 | 0.918 | 0.922 | 1.1 | 5.5-5.3 |
| 8.5 | 0.905 | 0.895 | 1 | - |
| 9 | 0.903 | 0.897 | 2.4 | 5.8-6 |
| 9.5 | 0.899 | 0.898 | 0.7 | - |
| 10 | 0.897 | 0.893 | 1.6 | 4.4-4.6 |

Moreover, the simulated radial dose function was quantitatively determined through fitting $5^{th}$ order polynomial function to the simulated values:

$$g(r) = a_0 + a_1 r + a_2 r^2 + a_3 r^3 + a_4 r^4 + a_5 r^5 \quad (12)$$

Where $a_0$, $a_1$, $a_2$, $a_3$, $a_4$ and $a_5$ coefficient of the radial dose function were obtained as 1.0142, 0.02236, $-1.3581 \times 10^{-1}$, $2.98 \times 10^{-2}$, $-3.76 \times 10^{-3}$ and $1.3569 \times 10^{-4}$, respectively.

Fig. 4 shows the comparison between the simulated and practical radial dose function of the source under study, and that of other commercial sources. According to Fig. 4, there is a great agreement between the simulated radial dose function of $^{192}$Ir-PDR and those reported for other commercial $^{192}$Ir sources.

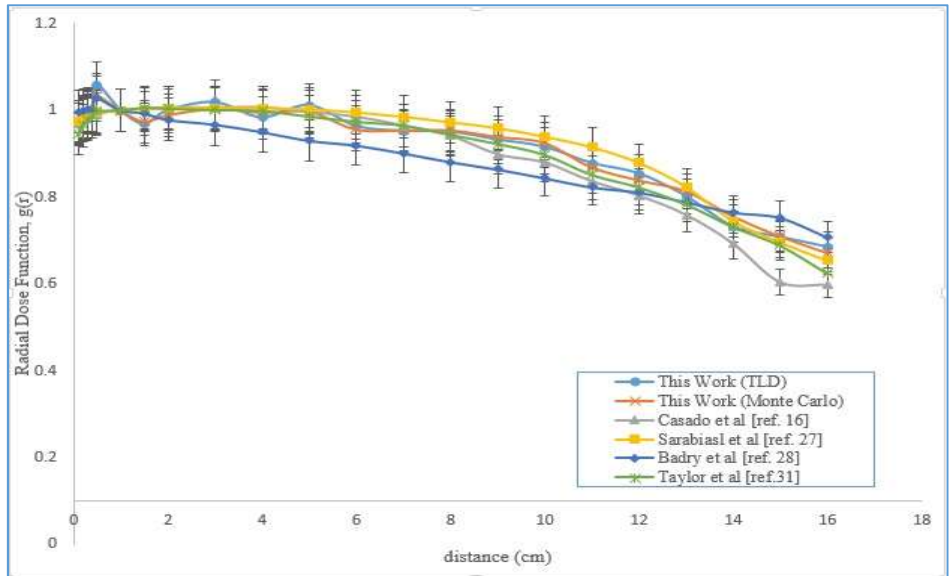

**Figure 4.** C Radial dose functions $g(r)$, as a function of radial distance $r$. The MC and measured calculations for the new PDR source in this work were compared with results from similar sources.

The anisotropy function of $^{192}$Ir-PDR was measured in water phantom using LiF TLD at angles ranging from 0°– 180° at 10° intervals relative to the source axis at radial distances of 2 cm, 3 cm, 5 cm, and 7 cm. Moreover, the anisotropy functions of this source were calculated in water phantom using Monte Carlo simulation technique at angles ranging from 0° – 180° at 5° intervals relative to the source axis. These results are reported in Table 5.

**Table 5.** TLD measured and Monte Carlo simulated anisotropy functions of $^{192}$Ir brachytherapy source. The average anisotropy function is seen in the last row of this table.

| F(r,θ) | | | | | | | | |
|---|---|---|---|---|---|---|---|---|
| | Measured with TLD (Plexiglass) | | | | Monte Carlo simulated (Water phantom) | | | |
| Angle θ Degree | r=2 cm | r=3 cm | r=5 cm | r=7 cm | r=2 cm | r=3 cm | r=5 cm | r=7 cm |
| 0 | 0.839 | 0.835 | 0.837 | 0.823 | 0.765 | 0.785 | 0.734 | 0.808 |
| 5 | - | | | | 0.796 | 0.862 | 0.814 | 0.876 |
| 10 | 0.888 | 0.931 | 0.888 | 0.952 | 0.828 | 0.933 | 0.837 | 0.942 |
| 15 | - | | | | 0.789 | 0.928 | 0.931 | 0.934 |
| 20 | 0.923 | 0.968 | 0.921 | 0.969 | 0.953 | 0.938 | 0.938 | 0.959 |
| 25 | - | | | | 0.945 | 0.954 | 0.944 | 0.963 |
| 30 | 0.968 | 0.987 | 0.980 | 0.968 | 0.926 | 0.970 | 0.974 | 0.975 |
| 35 | - | | | | 0.884 | 0.963 | 0.956 | 0.981 |
| 40 | 0.973 | 0.996 | 0.959 | 0.984 | 0.953 | 0.972 | 0.962 | 0.982 |
| 45 | - | | | | 1.002 | 0.977 | 0.978 | 0.995 |
| 50 | 0.988 | 1.013 | 0.973 | 1.008 | 1.004 | 0.982 | 0.984 | 1.012 |
| 55 | - | | | | 1.014 | 0.987 | 0.995 | 1.003 |
| 60 | 1.001 | 1.009 | 1.001 | 1.006 | 1.025 | 0.998 | 1.006 | 1.018 |
| 65 | - | | | | 1.043 | 1.009 | 1.003 | 1.011 |
| 70 | 0.996 | 0.993 | 0.992 | 1.006 | 1.012 | 1.015 | 1.018 | 1.031 |
| 75 | - | | | | 1.033 | 1.008 | 1.012 | 1.017 |
| 80 | 1.008 | 0.999 | 0.981 | 0.984 | 1.028 | 1.026 | 1.029 | 1.026 |
| 85 | - | | | | 1.013 | 1.021 | 1.019 | 1.024 |
| 90 | 1 | 1 | 1 | 1 | 1 | 1 | 1 | 1 |
| $\Phi_{an}$ | 0.952 | 0.969 | 0.948 | 0.966 | 0.945 | 0.962 | 0.951 | 0.975 |
| $\Phi_{an\ (average)}$ | 0.958±5% | | | | 0.956±3% | | | |

Simulated and measured 2D anisotropy functions of $^{192}$Ir-PDR at two radial distances of 2 and 5 cm have been compared in Fig. 5; as it is clear in this Figure, the results of simulation are in accordance with the experimentally measured data. The uncertainty of the measured data is 4.5% and uncertainties of the calculated values are within 3.2%. This fact shows that the source geometry is accurately modeled by the Monte Carlo code. Fig. 5 shows an excellent agreement between the measured and calculated F(r,θ) in the water phantom, which further supports the results of the Monte Carlo simulations obtained herein.

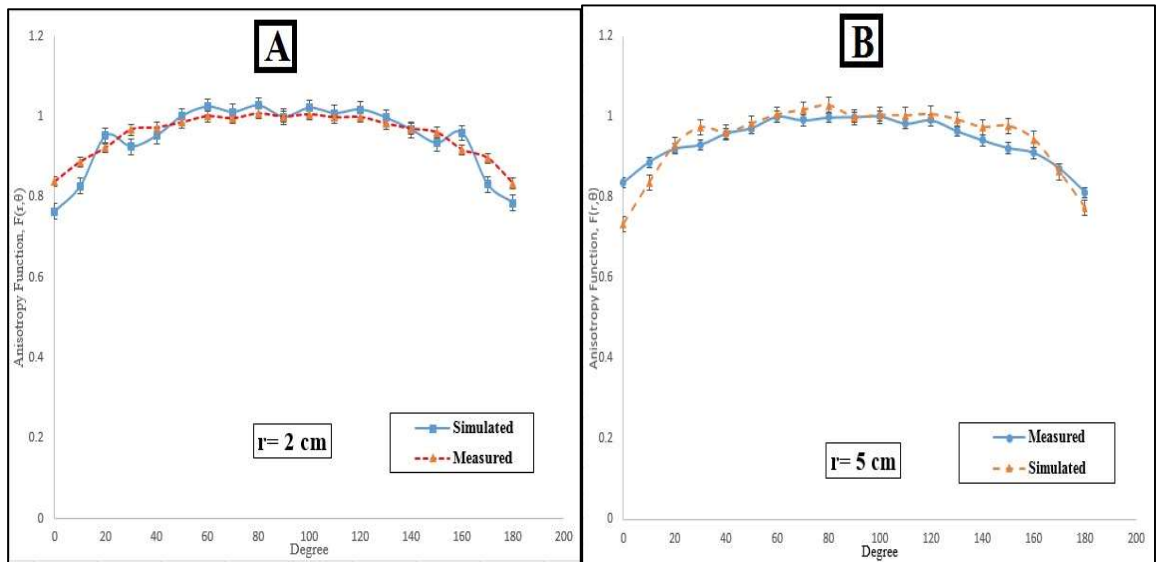

**Figure 5.** Comparison of anisotropy function for radial distance A) r = 2 cm B) r = 5 cm.

In Figs. 6a–c values of the function F(r,θ) for 192Ir seed source at distances of 2, 3 and 5 cm are compared with those reported by Taylor et al [27], Casado et al [16], Sarabiasl et al [24] and Badry et al [25]. Good agreement is apparent between the F(r,θ) obtained in this work and the data published by others. The relatively small differences between the present values and previous findings are suggested to be caused by the differences between phantom materials, self-filtration, oblique filtration of photons through the encapsulating material and active lengths of the sources. Fluctuations of values of F(r,θ), at symmetric angles may be due to non-uniformity of encapsulation of the sources. The results for the anisotropy function are plotted in Fig. 5a-d.

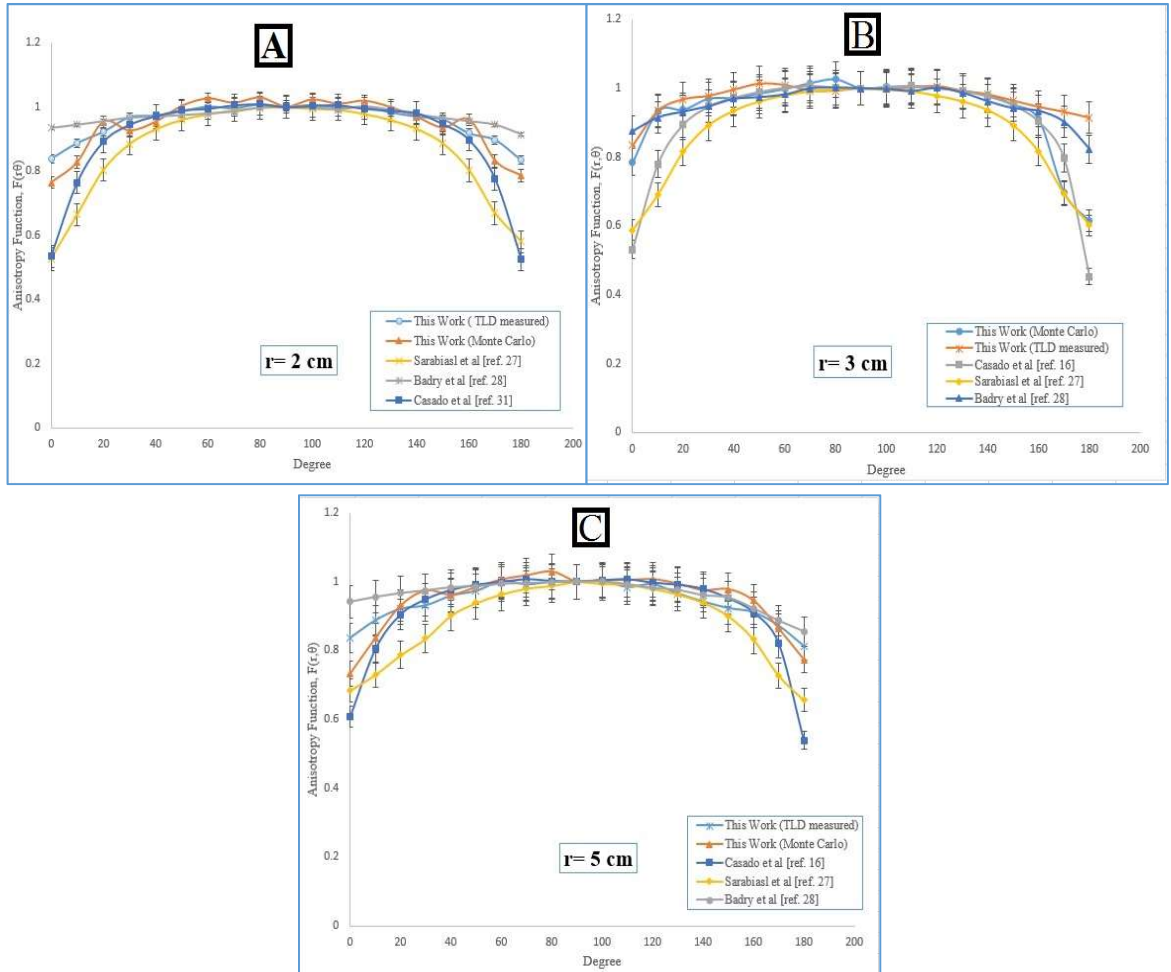

**Figure 6.** Comparison of anisotropy functions at different distances from the source center with respect to those of other sources. A) r = 2 cm, B) r = 3 cm, C) r = 5 cm.

## 4. Discussion

Recently, the use of brachytherapy sources as a permanent and temporary planting for the treatment of cancer is widespread. In this study, The TG-43(U1) [5] dosimetric characteristics (i.e. dose rate constant, radial dose function, and anisotropy function) of a newly designed $^{192}$Ir-PDR source have been measured in Plexiglass using LiF TLD chips and calculated in water phantom using the Monte Carlo simulation technique. The results of these determinations are presented. The new source design is different with respect to that of its predecessor in the length of the active core (L=3 mm for the new $^{192}$Ir-PDR source). The dose rate constant, radial dose functions, anisotropy functions, and dose rate distributions were calculated for the new source and the results were compared to the corresponding calculations for the other Source and microSelectron HDR,

PDR, and LDR sources. Moreover, accurate Monte Carlo modeling of the absorbed dose and underlying radiation interactions within the body (either in internal or external radiation planning [41], or imaging systems [42, 43]) would enable the creation of the comprehensive training dataset for training of the machine learning algorithms in order to reduce the dose calculation time [6, 44, 45]. Besides, machine learning algorithms in conjunction with the Monte Carlo modeling would establish synergies for precise and patient-specific measurement of underlying parameters [46, 47] and/or correction factors [48-50].

The dose rate constant, $\Lambda$, was found to be 1.131±0.007 using LiF TLDs in the Plexiglass phantom. This value was in good agreement with the Monte Carlo simulation in water phantom which was 1.173±0.005. The agreement between the measured and calculated dose rate constant indicated that correct source geometry has been used in the simulation. The results of comparing the dose rate constant with other similar sources indicate an appropriate agreement in between. These results are displayed in Table 3. The significant differences between the dose rate distributions calculated around the investigated sources, observed at short radial distances and/or along their longitudinal axes, are due to the different geometric characteristics of these sources.

From equation (8), the radial dose function was estimated using the output files of Monte Carlo simulations and measured practical with TLD (the values the geometry function calculated with equation (7)). The dose distribution was normalized at 1 cm from the source center (reference point) as described in the equation (8). The values of the radial dose functions $g_L(r)$ were tabulated (Table 4) and presented in Fig. 4 for the distances ranging from 0.5 cm to 16 cm. To validate the results of this simulations and measured practical of the radial dose function, a comparison was made with the published values of Taylor et al [25], Casado et al [16], Sarabiasl et al [24] and Badry et al [25]. The difference in the RD% , in comparison with the reference data and the data Sarabiasl et al [24] presented, was calculated to be between 0.1 to 5.8 percent and 0.1 to 6 percent, respectively. For the radial dose function, there is an increase in the statistical errors for Monte Carlo calculations and measured practical with TLD with an increase in the radial distance; however, the number of histories remains constant. Therefore, the maximum uncertainly was 6% (for r= 9 cm point).

The anisotropy function, $F(r,\theta)$, of the New $^{192}$Ir-PDR source was measured and calculated at 10° and 5° increments, respectively, relative to the source axis (Table 5), at distances of 2, 3, 5 and 7 cm using LiF TLD and Monte Carlo simulation. As shown in Fig. 5a-c, the anisotropy function calculated through the use of measurement and simulation shows an appropriate relative agreement with other similar sources. The anisotropy factors, $\varphi_{an}(r)$, and anisotropy constant, $\varphi_{an}$, were determined using Eq. (11). The result shows an excellent agreement between the measured (0.958±5%) and calculated (0.956±3%) anisotropy constant of the new 192Ir-PDR source in Plexiglass and water phantom, respectively. Fig. 5 (a), Fig. 5 (b), and Fig. 5 (c) for distances 2, 3 and 5 cm, respectively, show the results of comparing the anisotropy function to other sources of

Casado et al [16], Sarabiasl et al [24], and Badry et al [25]. The relative agreement with the maximum error of 6% is evident.

Lastly, this is an initial dosimetric characterization of the new $^{192}$Ir-PDR source that was performed experimentally and theoretically based on TG-43(U1) recommendations [5]. This information is available in tabulated and graphical format, suitable for the most commonly available treatment planning systems. The results of the Monte Carlo simulation with regard to the practical and laboratory validation performed in different researches are considered as an important recommendation. The findings of this study confirmed that the dosimetric characteristics of this new source are comparable to those of other commercially available sources, which are practically being used in clinical radiation oncology applications. Based on this fact, it can be concluded that this new source has the potential of a clinical brachytherapy source. Furthermore, there was a good agreement between the results of Monte Carlo simulation and our TLD measurements. This fact demonstrates that the simulated dosimetric parameters can be employed to obtain the dose distribution imposed by the new brachytherapy source.

## 5. Conclusion

In conclusion, the dosimetric characteristics of the new $^{192}$Ir-PDR source were experimentally characterized based on TG-43(U1) recommendation. Close agreement was observed between the results of Monte Carlo simulation and our TLD measurements. The finding of this study confirmed that the dosimetric characteristics of this new source are comparable to those of other commercially available sources indicating that this new source has the potential to be a clinical brachytherapy source.